\documentclass[9pt,oneside]{amsart}
\usepackage{url}
\usepackage{cancel}
\usepackage{xspace}
\usepackage{graphicx}
\usepackage{multicol,caption}
\usepackage{subfig}
\usepackage{amsmath}
\usepackage{amssymb}
\usepackage[a4paper,width=170mm,top=18mm,bottom=22mm,includeheadfoot]{geometry}
\usepackage{textcomp}
\usepackage{booktabs}
\usepackage{array}
\usepackage{verbatim}
\usepackage{caption}
\usepackage{cite}
\usepackage{float}
\usepackage{pdflscape}
\usepackage{mathtools}
\usepackage[pdftex,usenames,dvipsnames]{xcolor}
\usepackage{afterpage}
\usepackage{tikz}
\usetikzlibrary{arrows,backgrounds,snakes}
\usepackage{multido}
\usepackage{color}
\usepackage{dependencies}

\title{Acurast: Decentralized Serverless Cloud \\  ~ \\{\smaller \textbf{0.3 - Jan 08, 2026, White Paper \\ ~}}}

\author{
  Christian Killer, Alessandro De Carli, Pascal Brun,\\ Amadeo Victor Charlé, Mike Godenzi, Simon Wehrli\\ \\
  Acurast Association, Baarerstrasse 43, 6300 Zug\\
$\langle$ first name $\rangle$ @acurast.com }

\begin{document}
\begin{abstract}
Centralized trust is ubiquitous in today's interconnected world, from computational resources to data storage and its underlying infrastructure. The monopolization of cloud computing resembles a feudalistic system, causing a loss of privacy and data ownership. 

Cloud Computing and the Internet in general face widely recognized challenges, such as \1 the centralization of trust in auxiliary systems (\eg centralized cloud providers), \2 the seamless and permissionless interoperability of fragmented ecosystems and \3 the effectiveness, verifiability, and confidentiality of the computation. Acurast is a decentralized serverless cloud that addresses all these shortcomings, following the call for a global-scale cloud founded on the principles of the open-source movement. 

In Acurast, a purpose-built orchestrator, a reputation engine, and an attestation service are enshrined in the consensus layer. Developers can off-load their computations and verify executions cryptographically. Furthermore, Acurast offers a modular execution layer, taking advantage of secure hardware and trusted execution environments, removing the trust required in third parties, and reducing them to cryptographic hardness assumptions. With this modular architecture, Acurast serves as a decentralized and serverless cloud, allowing confidential and verifiable compute backed by the hardware of security and performance mobile devices. 

\end{abstract}

\maketitle

\setlength{\columnsep}{20pt}
\begin{multicols}{2}

\section{Introduction}
\label{sec:introduction}
The broad adoption of cloud computing has resulted in an Internet landscape that is strikingly similar to feudalism~\cite{feigenbaum2014}. The dominance of a few monopolistic players created a scenario where trust in cloud operators is a requirement, not just an option. Technology critics even argued that the \enquote{Lords of the Cloud} are wreaking havoc on the world economy~\cite{lanier2014who}, causing loss of privacy, data ownership, and ease of mass surveillance~\cite{feigenbaum2014}. The only way forward is not simply to \textit{ distribute} the Internet's architecture, but \textit{decentralize} it at large. A suitable path forward is to develop a global-scale cloud founded on open source principles, therefore aligning with the public interest~\cite{feigenbaum2014}.

\epigraph{Trust is our only option. In this system, we have no control over the security provided by our feudal lords. We don’t know what sort of security methods they’re using, or how they’re configured.}{Bruce Schneier, 2012 \cite{schneier2016cloudfeudalism}}

The fundamental prerequisites for a global-scale cloud network founded on open-source principles are: \1 global-scale connectivity, \2 widespread access to computation (\eg smartphones), and \3 societal acceptance of permissionless systems and their adoption. 

First, the rapid spread of smartphones has profoundly influenced the modern digital landscape. As of mid-2023, nearly 96 percent of the global digital population had accessed the Internet using a mobile device. Furthermore, improved efficiency are exemplified by ARM-based architectures. Reduced Instruction Set Computing (RISC) forms the foundation of the Advanced RISC Machine (ARM) architecture, providing superior energy efficiency compared to traditional x86 architectures. 

The proliferation of smartphones has also influenced laptops and desktops to adopt ARM-based instruction set architectures, even expanding to commercial environments, where ARM chips are increasingly employed in cloud servers~\cite{tamid2024riseOfArm}. 

In parallel to these developments, society's understanding of cryptoeconomic systems has evolved. Cryptographic tools have been developed in an open source way, shaping society's willingness to deploy cryptoeconomic systems~\cite{tarasiewicz2015}. The rapid development and adoption of public blockchains has driven forward real-world use cases. These use cases demonstrate the long-term impact and relevance to key elements of society's fabric that create resilient and \textit{ unstoppable} systems~\cite{graypaper}. 

A key property of decentralized protocols is the verifiability of the computation. In blockchain-based systems, the consensus layer ensures that state changes are verified and the consensus rules are enforced and followed. Relying on distributed consensus guarantees high security; however, in practice, it means that every participating node has to perform the computation, also. 

The blockchain trilemma is a guiding principle in the design and implementation of blockchains. However, the trilemma ignores two crucial factors: the first is the effectiveness of the computation. Thus, the blockchain quadrilemma~\cite{mogavero2021quadrilemma} further extends this notion, considering the ability \emph{computational effectiveness}, \ie to run complex decentralized computations at affordable costs, which poses a fundamental problem in blockchains today. 

Approaches addressing these issues often ignore crucial factors, such as the hardware layer \ie the \textit{trustless} decentralization of its physical execution layers, leading to centralization of crucial auxiliary systems (\eg cross-chain bridges or sequencers). A recent study estimates that approximately \$2 billion in cryptocurrency has been stolen in 13 different cross-chain bridge hacks, representing 69\% of the total stolen funds alone in 2022~\cite{Chainalysis2022}. Ironically, many of these auxiliary systems are run on public clouds, and by now public cloud providers are even actively providing blockchain tools~\cite{coindesk2023web3amazon}.  

Furthermore, a major strength of blockchains can be considered as their main weakness: the public and transparent availability of all events for anyone to read~\cite{bowe2020zexe}. Thus, the second extension of the blockchain trilemma is that confidentiality is required for certain computations, and the public nature of blockchains negatively impacts the relevant execution. Therefore, we extend the quadrilemma with a nuanced notion of \emph{computational effectiveness and confidentiality}.

\epigraph{In words from history, let us speak no more of faith in man, but bind him down from mischief by the chains of cryptography.}{Edward Snowden~\cite{greenwald2014no}}

Acurast is a novel decentralized serverless cloud that addresses the shortcomings of today's computing infrastructure. With its novel zero trust approach, Acurast \textit{trustlessly} provides a confidential and verifiable compute infrastructure. Acurast follows a modular separation of the consensus, execution, and application layer. The Acurast orchestrator is embedded in the consensus layer and contains a purpose-built reputation engine to ensure reliability and incentivize honest behavior. Furthermore, the consensus layer contains an Attestation service, ascertaining that the computation is end-to-end verifiable.

The foundation of the Acurast execution layer is made up of individual compute units (\ie smartphones). Smartphones are extremely energy efficient and yield better performance-per-watt compared to traditional server hardware~\cite{vonderassen2024depin}. Furthermore, ARM-based Systems-on-a-Chip (SoC) not only outperform traditional hardware, but also offer more secure hardware.  

For instance, on Android-based smartphones, the Acurast Trusted Execution Environments (TEE) is bootstrapped by fully locking down the OS. Then, the TEE signs and publishes the public key on the Attestation service, allowing anyone to verify the authenticity of the TEE and hardware. The deployment of TEE reduces the trust assumptions to \1 cryptographic hardness assumptions and \2 a single root of trust (\ie secure hardware with key attestation and an external coprocessor). 

Acurast lays the foundation for a confidential, verifiable, and decentralized cloud. Potential use cases range from computation (\eg Federated Learning), or data  blockchain infrastructure (\eg Zero Knowledge-as-a-Service, On-Chain Automation, Price feeds) to distributed trust in Federated Learning (FL)~\cite{lo2023distributed}, native cross-chain DeFi capabilities, and privacy mixing~\cite{diaz2021nym, theis2022mixclaves}. 

The remainder of this work is organized as follows. An end-to-end deployment is detailed in Section~\ref{sec:end2end}. The Acurast architecture is described in Section~\ref{sec:architecture}, which contains an overview of the high-level architecture and its different layers. The zero trust execution layer is described in Section~\ref{sec:execution_layer}. Section~\ref{sec:consensus_layer} discusses the consensus layer. Finally, Section~\ref{sec:application_layer} describes different areas of application of Acurast, and Section~\ref{sec:summary} summarizes.

\begin{figure*}[t]
\centering
\includegraphics[width=1\textwidth]{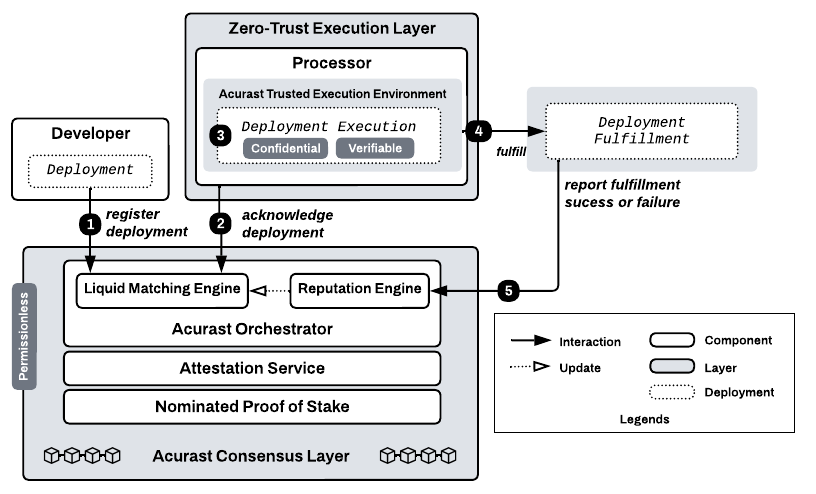}
\captionof{figure}{End-to-End Zero Trust Deployment Execution}
\label{fig:endtoend}
\end{figure*}
\section{End-to-End Zero Trust Deployment}
\label{sec:end2end}
Acurast introduces a paradigm shift in \emph{verifiable and confidential} computation, advancing the way decentralized clouds are operated, and applications are developed and deployed.
To emphasize the inner workings of Acurast, the following description follows a \texttt{deployment} from its prerequisite steps, its inception to the final completion (\cf Fig.~\ref{fig:endtoend}).

\textbf{(0) Processor Attestation} This step is a prerequisite to bootstrapping the Acurast Trusted Execution Envirnoment (TEE) of the individual processors. As a first step, the processor is generating an asymmetric key pair $(\pk,\sk) \sample \kgen$, whereas the private key of that pair is remaining in the dedicated secure element (\ie coprocessor, depending on the terminology of the hardware SoC). For example, in novel Google phones, the Titan M2 chip's dedicated coprocessor is used~\cite{Rossi2021titan}, whereas on Snapdragon Systems-on-a-Chip (SoC), the Qualcomm Secure Execution Environment (QSEE) is used). To conclude this step, $\pk$ is signed and sent to the Attestation service, where the trust root of the respective TEE can be verified (\eg Google as a trust root for Titan M2)

\textbf{(1) \texttt{Deployment} Registration:} As a first step, developers define their \texttt{deployment} details. For example, what kind of Runtime should be used, \eg Node.js. Also, whether or not the application contains a Web3 destination \ie at what destination the \texttt{deployment} should be \emph{settled}. In other words, on which protocol the \texttt{deployment} \texttt{output} should be persisted (\eg Bitcoin Mainnet), or whether the app is settled by calling a REST API (\eg web crawling). After that, the developer can select \emph{ready-to-deploy} templates, which can be adapted to the developer's needs, or a custom \texttt{deployment} can be defined. 

Depending on fidelity of integration of the destination ecosystem with Acurast, the pre-payments for gas fees and rewards are settled in the native currency the developer prefers (\eg native \texttt{TEZ} for Tezos or \texttt{ETH} for Ethereum) or in native Acurast \texttt{ACU} token. 

Next, the developer must state on which processors the \texttt{deployment} should be executed, either \textit{(a)} on personal processors, or \textit{(b)} on selected, known processors (\eg known trusted entities), or \textit{(c)} on public processors. For \textit{(a)}, a processor reward is not a hard requirement, since it is a permissioned setting. For \textit{(b)} a reward is optional, and for \textit{(c)} the liquid matching engine and the Acurast orchestrator will match appropriate processor resources with developers' \texttt{deployments}. 

In addition, more details of the \texttt{deployment} need to be declared, such as  \emph{scheduling} parameters, including the start time, end time, the interval between executions, as well as the duration in milliseconds and the maximum start delay in milliseconds. Furthermore, specific resource management parameters, such as memory usage, network requests, and storage requirements of the \texttt{deployment} need to be declared. Finally, the reward for the execution of the deployment should be declared, as well as the minimum reputation (only applies to \textit{(c)} public processors). Then the \texttt{deployment} will be persisted on the Acurast Consensus Layer and reaches \texttt{OPEN} state (\cf Fig.~\ref{fig:job_states}).

\textbf{(2) \texttt{Deployment} Acknowledgment:} Second, the processor acknowledges the \texttt{deployment} and fetches the details from the Acurast network. Depending on the fulfillment definition of the respective \texttt{deployment}, the Merkle root of the \texttt{deployment} with proof of assignment is persisted on the target destination (\eg on a different target chain). Now the \texttt{deployment} reaches the \texttt{MATCHED} state, and no other processors will attempt to acknowledge it. 

A prerequisite for assigning the deployment to the processor is that the processor can execute the \texttt{deployment} in full, following the \textit{all-or-nothing} principle (\cf Sec.~\ref{subsec:liquid_matching} for more details). Since \texttt{deployments} can have different scheduling configurations (\eg on demand, every minute, etc.). Therefore, if the processor acknowledges that all slots can be adhered to, the \texttt{deployment} reaches the \texttt{ASSIGNED} state.

\textbf{(3) \texttt{Deployment} Execution:} Next, the \texttt{deployment\_script} is executed in the processor runtime. In the illustrated example of Fig.~\ref{fig:endtoend}, the execution is performed inside of the Acurast Trusted Execution Environment (TEE) (\cf Sec. \ref{subsec:atee}), \ie because \emph{confidentiality} is ascertained by secure hardware \eg an isolated and external coprocessor (\eg Google's Titan Chip~\cite{Rossi2021titan}).

\textbf{(4) \texttt{Deployment} Fulfillment:} Once the \texttt{deployment} execution is completed, the output is delivered to the declared destination, which could be a REST-API, another AI model, or even another Web3 system (\eg Tezos, Ethereum). In case of a cross-chain transaction, the processor settles the gas fees on the destination chain, since the developer has locked the necessary reward and gas fee amount up front when registering the \texttt{deployment}.

\textbf{(5) \texttt{Deployment} Reporting:} After completion, the processor reports back to the Acurast Consensus Layer, more specifically to the reputation engine. If fulfillment was successful, the report contains a transaction hash of the target chain containing the fulfillment transaction. In case of failure, the report contains error messages. Finally, the \texttt{deployment} is now in \texttt{DONE} state.

To ensure the reliability of the Acurast protocol, the reputation engine is continuously fed with reliability metrics, for instance, right after \texttt{deployment} completion or failure. Furthermore, the attestation service is updated if a certificate is revoked (\eg a certain hardware module is compromised or the attestation expired).

\begin{Figure}
\centering
\includegraphics[width=1\textwidth]{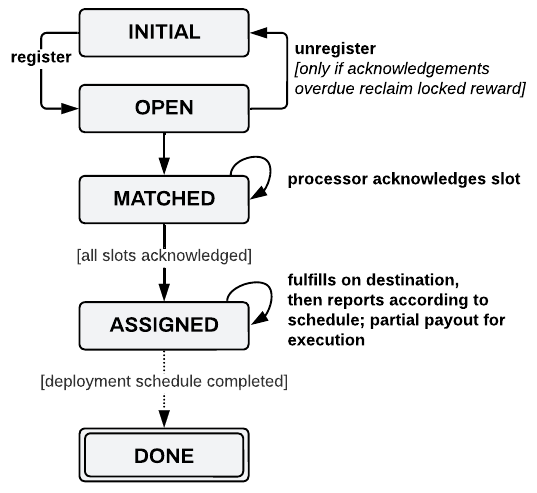}
\captionof{figure}{States of a deployment.}
\label{fig:job_states}
\end{Figure}

\section{Acurast Architecture}
\label{sec:architecture}
Acurast separates the consensus, execution, and application layer (\cf Fig.~\ref{fig:layers}). The consensus layer is the permissionless foundation of Acurast and is built on an adaptation of the Nominated Proof-of-Stake (NPoS) algorithm~\cite{burdges2020overview} (\cf Sec.~\ref{sec:consensus_layer}). The Acurast Orchestrator is a vital component of the consensus layer, matching consumer's \texttt{deployments} to processors, as previously outlined in the End-to-End flow (\cf Sec.~\ref{sec:end2end}). 

The second core part of the consensus layer is the reputation engine (\cf Sec.~\ref{subsec:reputation_engine}), which assures that the reputation scores of processors are correctly updated and incentivize honest behavior. A third key element is the attestation service which is responsible for the cryptographically verifiable attestations of the Acurast TEEs. The service is responsible for storing signed public keys, or revocations thereof. 

The Zero Trust Execution Layer (\cf Sec.~\ref{sec:execution_layer}) has major components. The first is composed of the execution environment, namely the Acurast TEE, \cf Sec.~\ref{subsec:atee}). The second key component are the Runtimes (\cf Sec.~\ref{subsec:runtimes}), which contains multiple modules that enable native interaction with different ecosystems. 

The third layer is the application layer, where applications composed of deployments are run (\cf Sec.~\ref{sec:application_layer}). Acurast infuses the development of a wide range of use cases that were previously not possible to implement in a permissionless and confidential manner. \textit{E.g.,} the trustless execution of market intelligence tasks (\ie web crawling), or the training and inference of an Artificial Intelligence (AI) models.  

In summary, Acurast's Zero Trust architecture transforms the way applications are designed and deployed. The modular nature allows for a truly decentralized, verifiable, and confidential compute platform, without introducing new trusted entities.

\begin{Figure}
\centering
\includegraphics[width=0.9\textwidth]{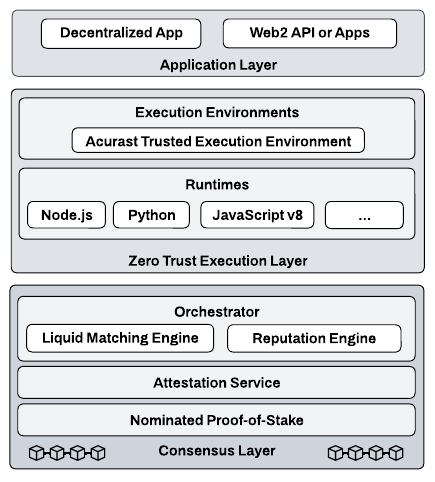}
\captionof{figure}{Acurast Architecture}
\label{fig:layers}
\end{Figure}

\section{Zero Trust Execution Layer}
\label{sec:execution_layer}
Acurast's execution layer is modular, allowing the flexible selection of runtimes according to the requirements of the use-case and the \texttt{deployment}, respectively. Decoupling the execution layer from the consensus and application layer allows the long-term evolution of the execution environment and its runtimes, avoiding dependency lock-ins. In addition, it ensures the highest level of service and confidentiality because security models can evolve iteratively with upgrades as novel threats emerge or new requirements arise. 

Acurast offers native and straightforward bootstrapping of permissioned consortia. Depending on the requirements, either \textit{(a)} consumers can directly leverage the Acurast orchestrator to select from a public pool of processors, or \textit{(b)} or use dedicated processors (\eg from trusted entities, or use consumer-supported self-service processors). This composability allows consumers to customize access control and define individual trust models depending on the \texttt{deployments} that are executed. 

The following sections outline the details of Acurast TEE (\cf~Sec.~\ref{subsec:atee}), then detailing the Acurast Runtimes (\cf~Sec.~\ref{subsec:runtimes}). Finally, verifiable computation and rationales on using smartphones are outlined. 

\subsection{Acurast Trusted Execution Environment}
\label{subsec:atee}
The Acurast Trusted Execution Environment (ATEE) is a generic approach to achieve a confidential execution layer while assuming a timely threat model, thus ensuring the highest possible level of security. The security guarantees achieved by secure hardware are generally highly divergent, from virtual processors to on-SoC processors, and finally, to the current bleeding edge of an \emph{external coprocessor}, which is a physically separated and independent chip, dedicated to only security-sensitive operations~\cite{Rossi2021titan}. 

Smartphones are among the most complex cases when it comes to information security. Their computing power has grown to the point of being almost indistinguishable from computers, they store the most valuable personal data and are used to carry out security-sensitive activities, which make them extremely attractive targets for attackers. With such a wide-ranging threat model and the fact that the vast computing base of a modern OS cannot be fully trusted, vendors have begun to use hardware to improve the security of their systems~\cite{Rossi2021titan}.

Thus, a highly secured ATEE instantiation is based on coprocessors provided by the Google Titan chip~\cite{android2023titanm2}. The Titan chip has not been compromised, unlike most secure hardware platforms. Furthermore the Titan M2 is a standalone processor with separate flash memory and minimal OS (often called microkernel). Crucially, these so-called coprocessors do not share memory or cache. Although high-reward bug bounties~\cite{reed2023bugbounty} and the highest zero-day vulnerability payouts~\cite{zerodium} do not \emph{guarantee} security, they are a solid indication of the security level achieved.

Usually, TEEs are created by integrating protection mechanisms directly into the processor or using dedicated external secure elements. However, both approaches only cover a narrow threat model, resulting in very limited security guarantees. For instance, enclaves nested in the application processor provide weak isolation and weak protection against side-channel attacks. Regardless of the approach used, TEEs often lack the ability to establish secure communication with peripherals, and most operating systems run inside TEEs do not provide state-of-the-art defense strategies, making them vulnerable to various attacks. Arguably, TEEs, such as Intel SGX~\cite{Schaik2020SGAxeHS, van2022sok} or ARM TrustZone~\cite{pinto2019arm}, implemented on the main application processor, are insecure, \emph{particularly} when considering side-channel attacks. For that reason, the security level of the ATEE is also dependent on the underlying hardware.

\subsection{Verifiable Computation}
A key property of decentralized computation protocols is the verifiability of the said computation. In blockchain-based systems, the consensus layer ensures that state changes are verified and the consensus rules are enforced and followed. Relying on distributed consensus guarantees high security; however, in practice, it means that every participating node has to perform the computation, also. For that reason, blockchains are impractical for use cases normally running on centralized clouds. 

\subsubsection{ZK-based Verifiable Compute}
A potential approach is presented by Zero-Knowledge Proof (ZKP) runtimes, as these allow for arbitrary computation (\eg Mina~\cite{mina}). While the verification of ZKPs is a low-effort computational task, ZKP generation is an intensive workload. Although ZKPs are becoming more efficient with time, the overhead of proof generation is most likely to remain in the foreseeable future~\cite{zkp-constant-overhead}. However, the complexity of ZKP implementations and its rapid evolution may introduce various vulnerabilities, undermining its very properties that it aims to introduce~\cite{zkp-security}.

\subsubsection{Secure Hardware-based Verifiable Compute}
Another approach to verifiable computation is the use of secure hardware. Acurast implements this approach in its ATEE (\cf~Sec.\ref{subsec:atee}, where secure hardware is leveraged to create an environment able to prove its integrity and allow for efficient and arbitrary verifiable computation.

\subsection{Runtimes}
\label{subsec:runtimes}
The Acurast execution layer natively offers multiple native \texttt{runtimes} that are continuously extended. \texttt{Runtimes} are APIs that are exposed to different apps running as \texttt{deployments}. For instance, runtimes include Node.js, JavaScript v8, and Python. Runtimes will be extended continuously.
\section{Permisionless Consensus Layer}
\label{sec:consensus_layer}

The Permissionless Consensus layer forms the base of the Acurast protocol and is based on a variant of the Nominated Proof-of-Stake (NPoS) algorithm~\cite{burdges2020overview}. Unlike traditional Proof-of-Stake (PoS) networks, there are \emph{validators} and \emph{nominators} in NPoS. Block validators verify transactions to be included in the next block, similar to traditional PoS block validators. The key difference is that instead of being randomly chosen, the \emph{validator} nodes are \emph{nominated} by another node. 

In Acurast's NPoS, an unlimited amount of token holders can participate as \emph{nominators}, backing a limited set of \emph{validators} with their stake. Having a limited set of \emph{validators} assures the long-term scalability of the consensus, allowing the increase of the maximum threshold through governance decisions~\cite{acurast_tokenomics}. An unlimited set of \emph{nominators} assure that higher value is at stake, assuring a high level of security. Due to its upgradable runtimes, consensus parameters are configurable by governance decisions, \eg The maximum number of \emph{validators}, and the minimum amount of stake for \emph{validators}. 

The Acurast NPoS system heavily leverages nominators to ensure network integrity. Nominators and validators have multiple aligned incentives. Nominators hold a financial stake in the system, which means that they could suffer a loss if a validator acts maliciously. In addition, nominators are financially rewarded for selecting a reliable and high-performance validator. Both nominators and validators have reputational stakes, with the credibility of nominators affected by their validator choices. Finally, the limited number of validator slots in NPoS structures creates a competitive environment, pushing nominators to select the most efficient validators, but also democratizes the process, making voting power essential.

NPoS has proven to be an efficient way to achieve high levels of security, scalability, and decentralization over time. Congruent to~\cite{burdges2020overview}, nominators share the rewards, or eventual slashings~\cite{acurast_tokenomics}, with the validators they nominated on a \emph{per-staked-\texttt{ACU}} basis. 

\subsection{Attestation Service}
\label{subsec:attestation_service}

The Attestation service is an integral part of the Acurast security mechanisms. Acurast uses secure hardware (\cf Sec.~\ref{subsec:atee}) to attest its processors hardware and the executable in a fully remote fashion, \ie a remote attestation is requested, which cryptographically links the binary hash of the executable $e$ (\ie app \texttt{deployment}) with the secret key $\sk_p$, remaining securely in the ATEE of the respective processor $p$. 

The respective key pair $(\pk_p,\sk_p)$ can only be accessed by that specific executable $e$, if $e$ is tampered with or deleted, and the access to $\sk$ is lost. The chain of trust is cryptographically verifiable from the executable $e$ to the original hardware manufacturer, allowing to form a verifiable execution environment inside the executable, which can further specify the preconditions for \eg allowing the executable to sign payloads. 

For example, a specific NodeJS \texttt{deployment} $d$ receives its individual key pair $(\sk_d,\pk_d)$ linked to the code of the \texttt{deployment} itself. So, if the NodeJS code is changed, access to the key pair $(\pk_d, \sk_d)$ is lost forever, and only the said deployment can effectively sign payloads with that $\sk$. This allows an external party to deduce that if a specific cryptographic signature is linked to a remotely attested key pair, this signature could only be a result of the execution of that specific set of instructions, offering an end-to-end verifiable execution environment. 

In the Attestation Service, devices advertise their compute resources along with their attestations. Only devices with valid attestations can then be matched with deployments, giving developers the security guarantees of an untampered execution environment. Furthermore, the attestations also include valuable information of the specific model of the hardware security module as described above, effectively allowing the developer to choose the level of security they require for their \texttt{deployment}.

\subsection{Acurast Orchestrator}
\label{subsec:acurast_orchestrator}
The Acurast orchestrator is a centerpiece of the consensus layer, combining the orchestration (\ie the scheduling of \texttt{deployments} and enabling the liquid matching) of the Processor's computational resources and developers. The orchestrator plays an essential role in the definition, agreement, and enforcement of value exchange. 

The orchestrator is where the liquid matching engine (\cf Sec. \ref{subsec:liquid_matching}) pairs the advertised processor resources with the defined requirements. The orchestrator natively supports various price-finding mechanisms (\eg auctions and advertisements), making DevEx highly accessible and seamless. 

Every agreement between processor and developer is specified in an entity called \texttt{deployment}. The \texttt{deployment} specifies \1 a set of instructions that are executed on the processor, \2 its scheduling parameters, and \3 settlement configuration (\ie where the output is further processed or persisted), and \4 finally, the pricing. 

\subsection{Compute Costs and Rewards}

When scheduling a deployment, developers define the compute cost for execution in native ACU tokens. This cost is fixed at submission time, enabling deterministic financial planning and accurate budgeting of computational expenses throughout the deployment's duration.

Critically, all compute costs are paid as gas (transaction) fees. There is no direct reward flow from developers to processors. Instead, processors earn rewards through two primary mechanisms that align their incentives with overall network health.

\subsubsection{Base Benchmark Rewards}

Providers receive rewards based on the benchmarks of their participating devices, paid from Acurast's token inflation. A total of 10\% of inflation is distributed per epoch (approximately every 1.5 hours) among all compute providers, independent of their participation in staked compute. This baseline structure ensures all processors contributing valid resources receive compensation, regardless of deployment selection during that epoch.

These rewards are split across the four benchmark metric pools, where each device competes based on its specifications. Higher-quality hardware receives proportionally greater rewards, incentivizing providers to contribute performant devices. \textit{C.f.,} Table~\ref{tab:benchmark_weights} in Subsection~\ref{subsec:stakedcompute}.

\subsubsection{Staking Rewards}

Compute providers participating in staked compute earn additional rewards by committing compute power and staking tokens. The majority (70\%) of Acurast's inflation is distributed as staking rewards every epoch, allocated based on hardware performance, stake size, and commitment duration.

The staking mechanism also serves as a security measure. Processors with staked tokens have financial incentive to behave honestly. Malicious or negligent behavior can result in slashing penalties, creating strong disincentives against attacks or unreliable service.

\subsubsection{Deployment Execution Bonus}
When a processor executes a deployment during an epoch, it receives a bonus weight on its benchmark metrics. This bonus increases the processor's scoring, and therefore its share of rewards, in both the Staked Compute Pool and the Compute Pool for that epoch.

This mechanism ensures processors are incentivized to actively execute deployments rather than passively holding stake. The bonus scales with deployment complexity, appropriately compensating processors handling more demanding workloads. 

\subsubsection{Inflation Distribution}
Acurast's annual token inflation is structured to support various stakeholders contributing to network operation:

\begin{itemize}
    \item \textbf{70\% $\rightarrow$ Staked Compute Pool:} Rewards processors providing reliable computational resources.
    \item \textbf{15\% $\rightarrow$ Treasury:} Funds ongoing development, ecosystem grants, and partnerships through governance-directed allocation.
    \item \textbf{10\% $\rightarrow$ Base Benchmark Rewards:} Compensates all providers based on hardware capabilities, regardless of current deployment activity.
    \item \textbf{5\% $\rightarrow$ Block Producers (Collators):} Compensates validators maintaining blockchain infrastructure and producing blocks.
\end{itemize}

This distribution model aligns incentives across all network participants while ensuring sustainable growth of the compute ecosystem.
\subsection{Scheduling Engine for Liquid Matching}
\label{subsec:liquid_matching}
\begin{figure*}
\centering
\begin{tikzpicture}[thick]
    \fill[fill=gray!20] (0,0) rectangle (6,1);
    \filldraw[fill=green!20] (2,0) rectangle (5,1);
    \node[above] at (1,1.7) {$\textsc{execution}_{i,1}$};
    \draw [thick] (0,0) to (0,1); 
    \node[below] at (0,0) {$\text{start}_i$};
    
    \node[below] at (2,0) {$s_i$};
    \node[below] at (5,0) {$e_i$};
    \draw[<->] (2.05,0.7) -- (4.95,0.7)
        node[pos=0.5,below]{$\text{duration}_i$};
    \draw[<->] (0.05, 0.7) -- (1.95,0.7)
        node[pos=0.5,below]{$\text{start\_delay}_i$};
    \draw[<->] (2.1, 1.3) -- (9.95,1.3)
        node[pos=0.5,above]{$\text{interval}_i$};

    \draw[>-<] (0, -2) -- (3,-2)
        node[pos=0.5,below]{$\text{max\_start\_delay}_i$};
    \draw[>-<] (2, -3) -- (5.3,-3)
        node[pos=0.5,below]{$\text{acceptance period for report}_{i, 1}$};
    \draw (2,-1.9) -- (2,-2.1);
    \draw [dotted, -] (0,-2) to (0,0);
    \draw [dotted, -] (2,-2) to (2,0);
    \draw [dotted, -] (2,-2) to (5,0);
    \draw [dotted, -] (3,-2) to (6,0);

    \node[above] at (9,1.7) {$\textsc{execution}_{i,2}$};
    \fill[fill=gray!20] (8,0) rectangle (14,1);
    \filldraw[fill=green!20] (10,0) rectangle (13,1);
    \draw [thick] (11,0) to (11,1); 
    \node[below] at (11,0) {$\text{end}_i$};
    
    \node[below] at (-1,0.7) {$\texttt{deployment}_i$};

    \draw (-1,-4) -- (15,-4);
    \draw[snake=ticks,segment length=1cm] (-1,-4) -- (15.1,-4);

\end{tikzpicture}
\caption{\texttt{deployment} Execution Scheduling Parameters} \label{fig:schedule}
\end{figure*}

Acurast developers can retrieve data (\ie use Acurast as an Oracle framework) and relay the retrieved information to Smart Contracts (or other Web2 targets). Although developers could register an individual \texttt{deployment} for each required query, Acurast provides developers with a Scheduling Engine that guarantees a certain level of service. 

Given multiple scheduled executions, denoted $\textsc{execution}_{i,k}$, as one $\texttt{deployment}_i$, this allows Acurast to plan ahead and ensure that all executions, repeated with $\text{interval}_i$ throughout a schedule, fit the processor's capacity (\textit{all-or-nothing} principle). Furthermore, to track the level of service provided by processors and to specify Service Level Agreements (SLA) with the developer at the time of matching $\texttt{deployments}$ with processors by feeding the reputation system with reliability metrics.

Additional challenges are imposed on the protocol to achieve reliable scheduling, aligning incentives of all stakeholders correctly so that developers are ensured a high level of service from processors, which would damage their reputation if scheduled executions are not fulfilled. Lastly, processors can obtain commitments to predictable earnings, while the pricing for all executions is agreed on before the start of a $\texttt{deployment}$.

To perform more $\texttt{deployments}$ with flexible schedules that suit the processor capacity, Acurast introduces \textit{planning flexibility} for each $\texttt{deployment}_i$, denoted by $\texttt{max\_start\_delay}_i$. The matchmaker chooses the actual delay and is denoted by $\texttt{start\_delay}_i$. Figure~\ref{fig:schedule} sketches the relationship between all parameters of $\texttt{deployment}_i$ and the period within which the fulfillment and subsequent reporting to Acurast are accepted. Outside the acceptance period for $\texttt{report}_{i, 1}$ a processor submission is not accepted, leading to a \emph{malus} in the reputation system.

\subsection{Reputation Engine}
\label{subsec:reputation_engine}
Traditionally, reputation systems are used to gauge the trustworthiness of system participants. In Acurast, the trustworthiness and reliability of processors is crucial for the network's execution layer. Although Acurast's zero trust architecture significantly reduces the required trust assumptions, it is still a crucial metric \ie  reputation scores in Acurast indicate the processor's capability to provide the execution layer infrastructure reliably.

The binary input of the reputation engine is automatically obtained from the system parameters~\cite{amadeo2023thesis}. More specifically, a processor's reputation is determined by both successful and unsuccessful \texttt{deployment} fulfillment. Its statistical foundation is rooted in the beta Probability Density Function (PDF)~(\cf Fig.~\ref{fig:reputation_beta_distribution}), commonly used to represent the \emph{a posteriori} probabilities of binary events. The beta PDF can be expressed in terms of the gamma function $\Gamma$ as follows~\cite{josang2002beta}: 

\begin{align}
f(p|\alpha, \beta) &= \frac{\Gamma(\alpha + \beta)}{\Gamma(\alpha)\Gamma(\beta)}p^{\alpha-1}(1-p)^{\beta-1} \\
&\qquad\text{where } 0 \leq p \leq 1,\; \alpha > 0,\; \beta > 0 \notag
\label{eq:gamma}
\end{align}

\begin{Figure}
\centering
\includegraphics[width=0.8\columnwidth]{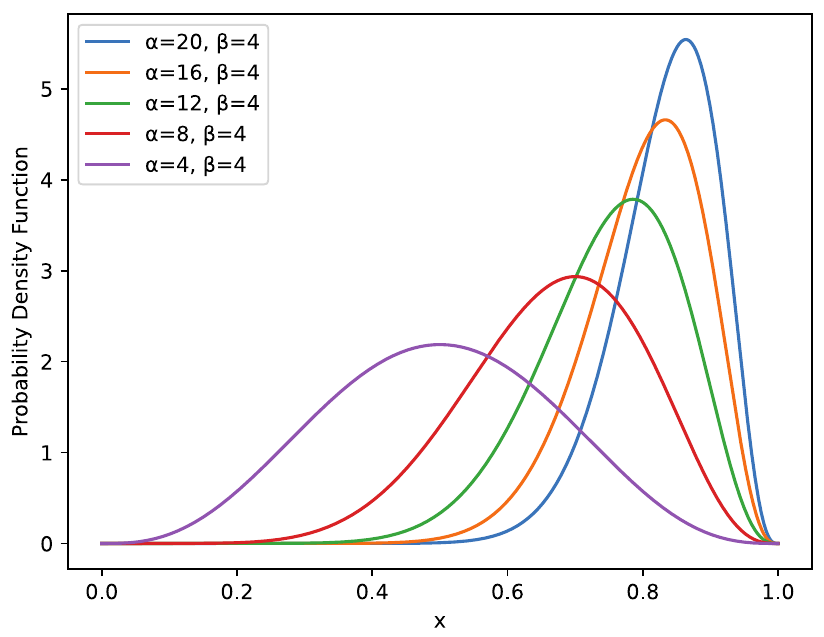}
\captionof{figure}{Beta Probability Density Functions}
\label{fig:reputation_beta_distribution}
\end{Figure}

Figure~\ref{fig:reputation_beta_distribution} displays PDFs of five different beta distributions with varying shape parameters $\alpha$ and $\beta$. Given the setting of binary events \{$A, B$\} Parameters $r$ and $s$ describe the observed frequency of occurrence of $A$ and $B$, respectively. Thus,  $\alpha$ and $\beta$ can be set as follows to model the PDF as a function of earlier observations:

\begin{equation}
\begin{aligned}
\alpha &= r + 1 \\
\beta &= s + 1
\end{aligned}
\bigg\}
\begin{aligned}
\; &r,s \geq 0 \\
\end{aligned}
\label{eq:2}
\end{equation}

Let $f_T$ represent the total number of favorable ratings observed given to entity $T$. Similarly, $g_T$ is defined as the total number of ratings given to the same entity~\cite{josang2002beta}. With the knowledge that the probability expectation value of the beta distribution is given as
$\mathbb{E}(p)=\frac{\alpha}{\alpha+\beta}$ and applying the definitions of Eq. \ref{eq:2}, thus the probability expectation value of our reputation function is defined as

\begin{equation}  
\label{eq:3}
Rep(f_T, g_T) = E(f(p | f_T, g_T)))  = \frac{f_T + 1}{f_T + g_T + 2}
\end{equation}

To account for context awareness, a weighting factor $w$ is introduced, which includes the \emph{transactional value} in the reputation model, thus preventing \emph{value imbalance} attacks~\cite{yao2012addressing}. In a reputation update, the weight of the underlying $\textsc{deployment}$ is determined as the ratio between the $\textsc{deployment}$ reward and the difference between the average $\texttt{deployment}$ reward (concerning the asset in which the reward is specified; excluding the reward of the $\texttt{deployment}$ for which the reputation update is invoked) and the $\texttt{deployment}$ reward. Let $n$ denote the number of assigned \texttt{deployments}. Then the weighting factor ($w_i$) for the reputation update for the $ith$ assigned $\texttt{deployment}$ is defined as 

\begin{equation}  
\label{eq:4}
w_i = 
\begin{cases}
    \;\;\;\;\;\;\;\;1\hfill, & \text{if } n=1 \\
    \frac{\phi_i}{\frac{1}{n-1}\sum\limits_{\substack{j=1 \\ j\neq i}}^{n}\phi_j + \phi_i}, & \text{if } n>1
\end{cases}
\end{equation}

By alternating honest and malicious behavior, an agent can strategically manipulate a reputation system~\cite{srivatsa2005trustguard}. To counteract this threat event, a discount factor $\lambda$~\cite{josang2002beta} is introduced to accurately represent the current reliability levels of the processors in real-time. A $\lambda$ value of $0$ implies that only the last reputation update is considered in the reputation scores, while a $\lambda$ value of $1$ does not have any discounting effects. In the presence of weights and $\lambda$, the order of reputation updates is crucial to calculate reputation scores. The upper summation limits refer to the most recent reputation update, and the lower limits refer to the oldest reputation update. Let $r_T$ denote the discounted and weighted favorable reputation updates. Similarly, let $s_T$ be the discounted and weighted total reputation updates. 

We can then express $r_T$ and $s_T$ as follows for $0<\lambda<1$: 
\begin{equation}
\label{eq:5}
\begin{aligned}
r_T = \sum_{i=1}^{n} r_{T,i} w_i \lambda^{n-i} \\
s_T = \sum_{i=1}^{n} s_{T,i} w_i\lambda^{n-i}
\end{aligned}
\end{equation}

Due to discounting of reputation updates, the maximum possible reputation score is less than in the naive case of the reputation function (\cf Equation \ref{eq:3}). 
Given the definition of $r_T$ \cf~Eq.\ref{eq:5}, the knowledge that $w_i < 1$ with the fact that the geometric sum $\sum\limits_{i=1}^{\infty} \lambda^i$ can be written as $\frac{1}{1-\lambda}$ for $\lambda < 1$, it follows that 
\begin{equation}
\label{eq:6}
\max_{\{r_T\}} = \frac{1}{1-\lambda}
\end{equation}

Let $\mu$ be defined as the maximum value of the reputation function, given $r_T, s_T, \lambda$ and setting $w=1$. It should be noted that in practice, $w$ is less than $1$ and, on average, $0.5$. Using the result of equation \ref{eq:6}, we can express $\mu$ as

\begin{equation}
\label{eq:7}
\mu = \frac{\frac{1}{1-\lambda} + 1}{\frac{1}{1-\lambda} + 2} < 1 \;\;\; \forall \lambda \in [0, 1)
\end{equation}

Scaling reputation scores using $\mu$, such that possible reputation scores are $\in (0,1)$. Equations \ref{eq:3}, \ref{eq:5} and \ref{eq:7} are combined to provide the following final version of our reputation formula:

\begin{equation}
Rep(r_T, s_T) = \frac{1}{\mu} \frac{\sum\limits_{i=1}^{n} {r_{T,i}} w_i \lambda^{n-i} + 1}{ \sum\limits_{i=1}^{n} {r_{T,i}} w_i \lambda^{n-i} + \sum\limits_{i=1}^{n} {s_{T,i}} w_i \lambda^{n-i} + 2}
\end{equation}

Under the assumption that arithmetic operations are executed in constant time and that the values $r_T$ and $s_T$ from Equation \ref{eq:4} are stored, the reputation algorithm has a time complexity of $\mathcal{O}(1)$.
Values \texttt{r} and \texttt{s} are persisted in the Substrate runtime storage for each processor. As such, the reputation scores are calculated from \texttt{r} and \texttt{s}, but are not stored separately. A developer can define the minimum reputation that a processor must have to be eligible to be assigned the specified $\texttt{deployment}$ when registering a $\texttt{deployment}$ on the market. 

\subsubsection{Evaluation}
The Acurast reputation system has been evaluated in \cite{amadeo2023thesis} using agent-based modeling (ABM) methods. The behavior of processors and developers was specified, and Acurast orchestrator interactions were captured in our model. Three distinct categories of processors were introduced: low, medium and high success rate groups with respect to their probability of fulfilling a standardized $\texttt{deployment}$. Evaluations followed a 200-step model of 300 processors and 900 consumers for ten iterations with values $p=0.8$, $p=0.9$ and $p=0.99$.

Initial orchestrator simulations were performed in the absence of the reputation system. Assuming a standardized $\texttt{deployment}$ with both agent types implementing a pricing strategy, the average \texttt{deployment} reward was 101.30 units. Although reputation scores were not considered in this scenario, reputation updates were tracked, resulting in Figure~\ref{fig:no-rep}. The visible difference in rewards is caused by the varying success rates of processor groups and the fact that rewards are only paid for successful fulfillment.

\begin{Figure}
\centering
\includegraphics[width=0.80\columnwidth]{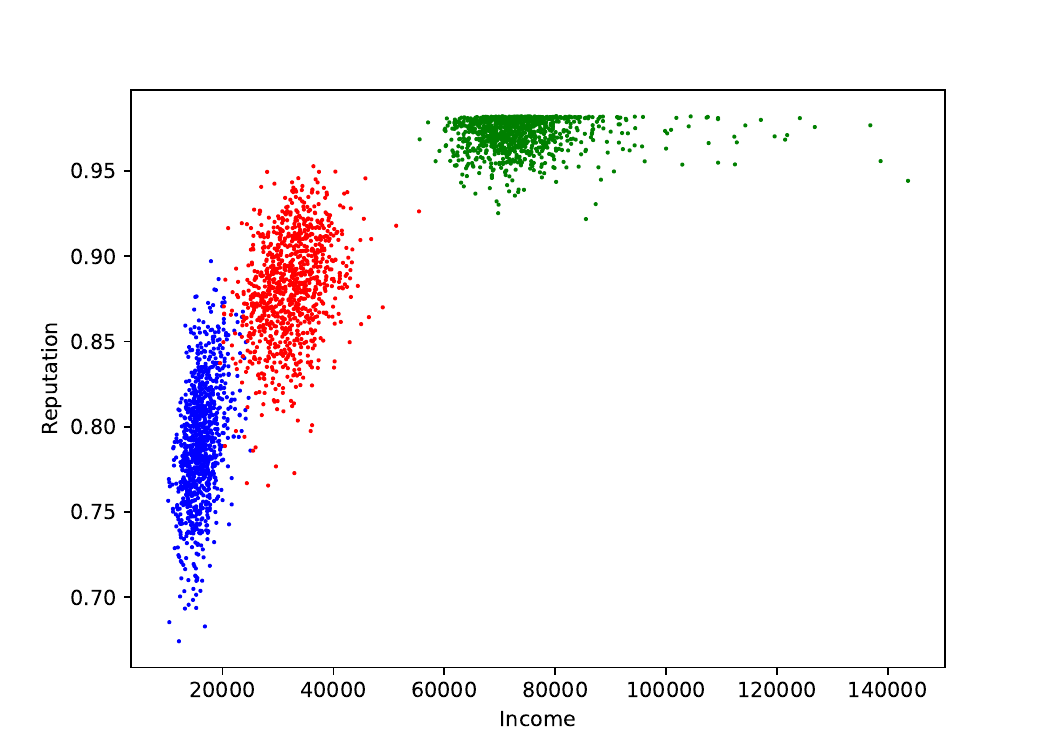}
\captionof{figure}{Simulation results obtained considering the reputation system in matching the reputation with the processor rewards. Outcomes of 10 iterations of a 200-step model with 3000 processor data points}
\label{fig:rep-80}
\end{Figure}

\begin{Figure}
\centering
\includegraphics[width=0.80\columnwidth]{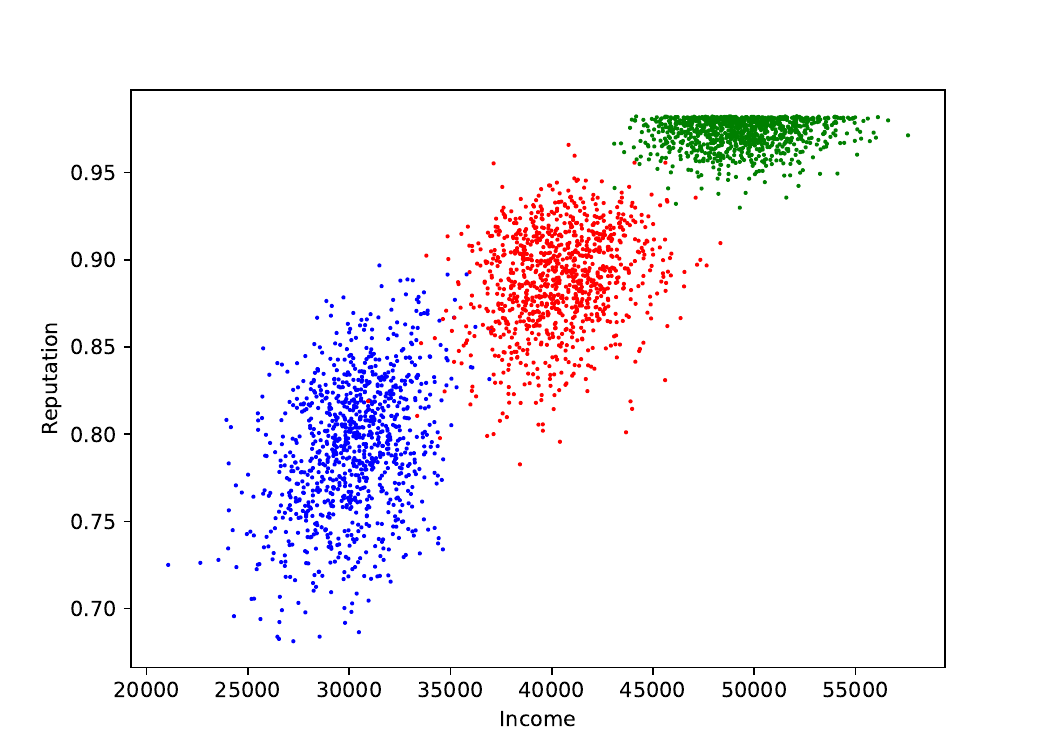}
\captionof{figure}{Simulation results obtained without the reputation system; reputation vs. processor rewards. Outcomes of 10 iterations of a 200-step model with 3000 processor data points}
\label{fig:no-rep}
\end{Figure}

The effect of the reputation engine on the matching is considered in another simulation scenario. Developers specify a minimum reputation in their \texttt{deployment} registrations with $p=0.5$, with the threshold value randomly drawn from a uniform distribution [0.8, 1]. 

Figure~\ref{fig:rep-80} visualizes the results. Evidently, there is a change in $\texttt{deployment}$ allocations with respect to the processor groups, in contrast to the previous scenario. Due to the fact that high-success processors are allocated a larger proportion of total assigned $\textsc{deployments}$, the average income for this group increases by 51.77\%. However, this occurs at the expense of the other groups, with the medium-success group booking 18.84\% less income while the low-success processor's income is diminished by ~46\%.

The Gini coefficient with respect to processor rewards is closely related to the percentage of \texttt{deployments} assigned to low-, medium-, and high-success processors. An important finding is that the presence of the reputation system significantly lowers the observed failure rate of allocated $\texttt{deployments}$ by 31.37\%, an effect that becomes even more pronounced if a higher percentage of developers demand a minimum reputation.

\section{Application Layer and Use Cases}
\label{sec:application_layer}
In the Internet of today, almost every application relies heavily on auxiliary systems. Whether external APIs are used for authentication, basic infrastructure (hosting), data availability, reliance, which can benefit from extending or replacing services and core elements with trustless \texttt{deployments}, essentially eliminating a host of threat events. The possibilities with Acurast are near-endless since today's centralized Internet is heavily centralized both logically and in terms of trust anchors. 

Acurast's execution layer transforms the way applications are designed and deployed, achieving an unparalleled Developer Experience (DevEx) by offering the Acurast Console~\cite{acurast_console} to developers, where a self-service model allows developers to integrate and develop their applications. Certain use cases can benefit tremendously from certain properties of Acurast, \eg geological distribution, confidentiality, verifiability. 

The following subsections outline potential use-cases that can be deployed through Acurast. 

\subsection{Market Intelligence}
Market intelligence involves collecting and analyzing data on a market, allowing businesses to make informed strategic decisions and stay competitive. Of all the Internet traffic in 2022, 47.4\% was automated traffic, also commonly referred to bots~\cite{Imperva}. Of that automated traffic, 30.2\% were bad bots, while good bots are also on the rise, accounting for 17.3\%. The percentage of human traffic continues its downward trend, from 57.7\% in 2021 to 52.6\% in 2022. 
Bots, in that context do not refer to volumetric Distributed Denial-of-Service attacks, but the bot activity on layer 7 of the OSI model. In general, good bots are important for various business models and applications (\eg market intelligence), since they are scraping data and feed models for decision making or business logic directly.

With Acurast, the scraping infrastructure can be fully decentralized logically and physically, leveraging the network of processor resources that confidentially execute these tasks, without leaking any data about the querying party. For example, when intelligence is gathered (\eg for investment or merger decisions), a large amount of data must be scraped confidentially. 

\subsection{Artifical Intelligence}
\label{subsection:ai}
The recent surge of Artificial Intelligence (AI) applications has led to increased research and development in these areas. Although the potential of these technologies is vast, the risks associated with the centralized deployment and privacy of data are crucial to evaluate carefully. In Acurast, the \emph{Singularity} module allows the execution of Artificial Intelligence (AI) in a decentralized and confidential fashion. \textit{E.g.,} Acurast enables Large Language Models (LLM) to be executed in a federated, privacy-preserving, and trustless way~\cite{Nguyen2021flsurvey}, \ie Distributed Llama~\cite{distributedLlama} Another example is Distributed LLama.

\subsection{IoT}
In general terms, the Internet of Things (IoT) refers to interconnected computing devices that form a network and monitor environmental variables (\eg health caren~\cite{baker2017IoT}). Often, due to its limited resources, heterogeneity, and lack of computing power, IoT faces many security and privacy challenges. Data is transferred between IoT devices without human intervention, making zero trust an essential aspect in network and trust management. Depending on the Acurast processor used, built-in Bluetooth modules or WiFi direct connections can be used to collect metrics or data and confidentially process the data.

\subsection{Zero-Knowledge Proof Applications}
Zero-Knowledge Proof protocols find vast application in blockchains~\cite{Xiaoqiang2021zkpsurvey}. Potential use cases range from anonymous voting systems~\cite{Killer2022ProvotuMN,Killer2022TNSM}, to secure and privacy-preserving digital assets exchange, secure remote biometric authentication, or Proof-of-Reserves~\cite{dutta2019proofofreserves}.

Acurast can be leveraged in multiple areas of ZKP applications, for instance, to offload high intensity computation in a zero trust manner~\cite{ni2023zksnarkgpu}, or to form sub-consortia of processors that can, for instance, mix and generate proofs that the mixing has been performed correctly. 

\subsubsection{Privacy-Preserving Mixing}
With ZKPs, privacy mixing can occur in a way that allows transactions to be validated without exposing the details of those transactions. However, the mixing is not limited to Web3 transactions and can be extended to other data sensitive to privacy (\eg metadata of internet traffic or files metadata). 

\subsection{Secure Multi-Party Computation}
Secure Multi-Party Computation (SMPC) is a cryptographic primitive that enables distributed parties to conduct joint computations without revealing their own private inputs and outputs to the computation~\cite{Wenliang2001}. For instance, doctors may query a database containing private information, or banks may invest in a fund that must satisfy both banks private constraints. Usually, one trusted entity must know the inputs from all the participants, however, if no Trusted Third Party (TTP) is available or suitable, privacy concerns are evident~\cite{Wenliang2001}. With Acurast, a processors can be selected for SMPC algorithms to execute \eg a permissionless poker game~\cite{kumaersan2016pokerbtc}.

\subsection{Blockchain Infrastructure}
Blockchain networks' rising adoption and complexity of blockchain networks has led to an increasing need for a reliable blockchain infrastructure. Novel incentive structures (\eg slashing in PoS) have intensified this further. It is crucial that this infrastructure is neither logically nor physically centralized because it would introduce new trust assumptions that undermine the permissionless nature of blockchains. For these services, Acurast can serve as a decentralized, serverless backend. 

\subsubsection{Incorruptible Sequencer}
A huge issue in public blockchains is Blockchain Extractable Value (BEV), and Miner Extractable Value (MEV), where DeFi users are at risk of being attacked~\cite{zhou2023sok} (\eg frontrunning and sandwich attacks~\cite{daian2019flash}). With Acurast, processors can serve as Sequencing-as-a-Service consortia~\cite{sharedsequencer}, ensuring that the order of transactions is deterministic and immune to external influence.

\subsubsection{Beyond Oracles: Serverless Applications}
Oracles and on-chain automation are key ingredients of blockchain infrastructure. While oracles enable external data to be imported into the blockchain, oracles mainly deal with data retrieval and validation, ensuring that accurate and reliable data is fed into smart contracts. However, on-chain automation has a broader scope, encompassing automated liquidity provision, periodic settlements, debt restructuring, yield harvesting, and much more. The emphasis here is on action and execution based on specific conditions.

\subsubsection{Native Cross-Chain DeFi}
Native cross-chain DeFi capabilities have been developed to allow seamless interactions and transactions between blockchains, creating a more inclusive and expansive financial ecosystem. Applying Account Abstraction allows for the design of accounts that can interact and integrate across various platforms and protocols, simplifying user experiences, and opening the door for innovative use cases. 

\subsubsection{Data Availability as-a-Service} (DAaaS) provides decentralized storage solutions to ensure data remain accessible and intact, fortifying the robustness of the entire decentralized ecosystem. 

\section{Tokenomics}
\label{sec:tokenomics}
The total supply of both \texttt{ACU} and \texttt{cACU} (Canary Acurast Tokens) is $1$ billion ($1.000.000.000$ tokens) subject to change based on the inflation and burn rate of the token. Details on token distribution and the Token Generation Events (TGE) are documented in~\cite{acurast_tokenomics,acurast_docs}. As a decentralized NPoS network, the protocol requires participants to stake (and nominate) to ensure that the protocol achieves its security and finality guarantees. Stakers are incentivized to participate by being rewarded with staking rewards that are derived from the inflation of the Acurast protocol. It is crucial to note that \texttt{ACU} tokens (\texttt{cACU} respectively) do not confer or represent any direct or indirect equity stake of the Acurast Association and/or the project or any other ownership right, share or equivalent right, voting right, or any other right, contractual or otherwise, to receive future profit shares, intellectual property rights, or any other form of participation.

\subsection{Staked Compute}
\label{subsec:stakedcompute}
  The continuity of service and the stability of compute supply are pivotal in long-term operations. Both the compute supply and demand sides require precise incentive alignment. For this reason, Acurast formulates a novel concept called \textit{staked compute}, allowing managers of compute resources to specify how long their compute will be available on the network. In addition, the manager assigns a certain stake to that commitment, ensuring a strong incentive for managers to adhere to their commitment without requiring a central authority. If the manager breaches their commitment, their stake will be reduced, with the penalty in relation to the duration of the breach.

  \subsubsection{Benchmark Metrics}
  In order to make compute measurable, Acurast introduces four standardized \textit{benchmark metrics} that quantify a device's computational capacity. Each metric is assigned a weight reflecting its importance to the network's computational requirements (\cf Tab.~\ref{tab:benchmark_weights})

  \begin{center}
  \begin{tabular}{lc}
  \hline
  \textbf{Metric} & \textbf{Weight} \\
  \hline
  CPU Single Core & 0.2307 \\
  CPU Multi Core & 0.2307 \\
  RAM & 0.4615 \\
  Storage & 0.0769 \\
  \hline
  \end{tabular}
  \captionof{table}{Benchmark metric weights}
  \label{tab:benchmark_weights}
  \end{center}

  These weights define how staking rewards are distributed across four separate \textit{benchmark metric pools}. For each unit of reward emitted, the distribution follows the weights above, with RAM being the most heavily weighted due to its critical role in application performance.

  \subsubsection{Current and Committed Compute}
  The \textit{current compute} represents the measured compute power across all of a provider's devices for a given epoch. These measurements are conducted automatically during each device heartbeat (every 30 minutes) and reported on-chain.

  A compute provider commits to maintaining a percentage of their current compute over the lifetime of their stake. The \textit{committed compute} is capped at 80\% of the measured current compute to prevent early slashing due to natural fluctuations. For each benchmark metric $p$, we denote the committed compute as $m_c^{(p)}$.

  \subsubsection{Staking Weight}
  The staking weight determines a participant's share of the reward pool. Let $s$ denote stake, $\tau$ denote cooldown period, and $\tau_{max}$ denote the maximum cooldown (approximately 3.68 years on Mainnet). For committer $c$:

  \begin{equation}
  \label{eq:committer_weight}
  W_c = s_c \times \frac{\tau_c}{\tau_{max}}
  \end{equation}

  The aggregate weight of delegators to committer $c$:

  \begin{equation}
  \label{eq:delegators_weight}
  W_D = \sum_{d}{ s_d \times \frac{\tau_d}{\tau_{max}} }
  \end{equation}

  \subsubsection{Stake Limits}
  To prevent disproportionate stake allocation relative to compute capacity, Acurast introduces a \textit{target weight per compute} for each benchmark metric pool $p$. Let $S$ denote total supply and $M^{(p)}$ denote total benchmarked metric:

  \begin{equation}
  \label{eq:target_weight}
  T^{(p)} = \frac{0.8 \times S}{M^{(p)}}
  \end{equation}

  This parameter expresses the ideal staking rate given the current compute offered across the network.

  \subsubsection{Reward Distribution}
  Rewards are calculated independently for each benchmark metric pool. The committer's score for pool $p$:

  \begin{equation}
  \label{eq:committer_score}
  \theta_c^{(p)} = \min\left(W_c + W_D, \; T^{(p)} \times m_c^{(p)}\right)
  \end{equation}

  Let $R^{(p)}$ denote total rewards for pool $p$. The reward share for committer $c$:

  \begin{equation}
  \label{eq:reward_share}
  r_c^{(p)} = R^{(p)} \times \frac{\theta_c^{(p)}}{\sum_{c}{\theta_c^{(p)}}}
  \end{equation}

  \subsubsection{Stake Delegations}
  Every compute manager can stake and accept stake delegations. The total \textit{commitment stake} consists of the manager's own stake plus all delegated stakes. To avoid centralization, Acurast enforces two limits:

  \begin{enumerate}
      \item The committer must provide at least $\frac{1}{10}$ of the total commitment stake (maximum $1{:}9$ ratio between own and delegated stake).
      \item Total commitment stake cannot exceed $T^{(p)} \times m_c^{(p)}$ for any pool $p$.
  \end{enumerate}

  Committers specify a \textit{delegation fee} ($\phi$) upon stake creation, applied when rewards are distributed. The fee can be decreased but never increased. For delegator $d$ with weight $W_d$, let $r_c$ denote total committer rewards and $\psi_d$ denote any slashing. The delegator reward:

  \begin{equation}
  \label{eq:delegator_reward}
  r_d = \frac{W_d \cdot (1 - \phi) \cdot r_c}{W_c + W_D} - \psi_d
  \end{equation}

  \subsubsection{Cooldown Period}
  When a committer or delegator initiates unstaking, a cooldown countdown begins. On Mainnet, cooldown periods range from 28 days to approximately 3.68 years. During cooldown:

  \begin{itemize}
      \item Reward weights are reduced to 50\%
      \item Slashing penalties remain at 100\%
      \item Full committed compute must be maintained
  \end{itemize}

  This mechanism protects the network from sudden losses of compute capacity.

  \subsubsection{Slashing}
  When current compute falls below committed compute in any metric during an epoch, slashing applies. Let $\rho = 0.003424657534\%$ be the maximum slash rate per epoch, $w^{(p)}$ the metric weight, and $\delta^{(p)}$ the shortfall ratio:

  \begin{equation}
  \label{eq:shortfall}
  \delta^{(p)} = \frac{m_{committed}^{(p)} - m_{current}^{(p)}}{m_{committed}^{(p)}}
  \end{equation}

  \begin{equation}
  \label{eq:slashing}
  \psi^{(p)} = s \times \rho \times w^{(p)} \times \delta^{(p)}
  \end{equation}

The total slash across all metrics cannot exceed $\rho \times s$ per epoch. Of slashed tokens, 10\% goes to the slasher; 90\% is burned. Delegators share slashing risk proportionally but may redelegate to a different committer at any time. Any updates to the Staked Compute mechanism will be documented in the official documentation~\cite{acurast_docs}.
\section{Summary}
\label{sec:summary}
Acurast is a novel decentralized serverless global-scale cloud built on open source. The execution layer is formed by a distributed backbone of mobile devices. With superior hardware and performance per watt, ARM-based SoCs offer a better alternative to centralized clouds. Acurast follows a modular architecture that separates the consensus, execution, and application layer. 

In the consensus layer, the orchestrator contains a liquid matching engine, which ensures that the supply (\ie processors) and the demand for computational resources (\ie consumers) match quickly. Furthermore, a custom-made reputation engine ensures reliability and incentivizes the genuine behavior of processors. Finally, the attestation service ensures that processors provide the Acurast TEE in a fully end-to-end verifiable manner. 

Acurast's execution layer offers the Acurast Trusted Execution Environment (ATEE), ascertaining verifiable and confidential compute, drastically reducing trust assumptions. Furthermore, various runtimes offer multiple native \texttt{runtimes} that are continuously extended, \eg Node.js, JavaScript v8, and Python. 

In summary, Acurast processors offer a flexible solution to run \texttt{deployments} tailored to application-specific requirements. Acurast presents a paradigm shift of verifiable compute, where trustless transfers across ecosystems are enabled through robust decentralized computational infrastructure, combining the benefits of confidentiality and censorship-resistance. The applicability of Acurast is broad, spanning from blockchain-specific utilities to distributed trust mechanisms in federated learning, and confidential computing. 

\section*{Acknowledgements}
The authors thank Rodrigo Quelhas, Matthias Tarasiewicz, Maximilian Fries, Ralph Pichler, Jake Blew, and Jan von der Assen for their input.

\bibliography{references}
\noindent \small{\\All links above were last accessed on January 8th, 2026.}
\addcontentsline{toc}{chapter}{References}

\bibliographystyle{IEEEtran.bst}

\end{multicols}

\end{document}